# Principles of Stochastic Computing: Fundamental Concepts and Applications

Sadra Rahimi Kari

*Abstract-* **The semiconductor and IC industry is facing with the issue of high energy consumption. In modern days computers and processing systems are designed based on Turing machine and Von Neumann's architecture. This architecture mainly focused on designing systems based on deterministic behaviors. To tackle energy consumption and reliability in systems, Stochastic Computing was introduced. In this research, we aim to review and study the principles behind stochastic computing and its implementation techniques. By utilizing stochastic computing, we can achieve higher energy efficiency and smaller area sizes in terms of designing arithmetic units. Also, we aim to popularize the affiliation of Stochastic systems in designing futuristic BLSI and Neuromorphic systems.**

## I. Introduction

The computer's perfectionist streak is coming to an end. Power consumption concerns are driving computing toward a design philosophy in which errors are either allowed to happen and ignored, or corrected only where necessary [1]. The traditional methods of correcting errors, meaning circling back to correct errors once they are identified, are power-consuming. Also, conventional design architectures dictate lower threshold voltages and higher working frequencies for designed switches. This combination leads to a faulty system, meaning by lowering the voltage and increasing clock frequency in designed Integrated circuits, we increase the error rate in our systems. The result is not ideal for conventional computing methods.

Stochastic computation, exploits the statistical nature of application-level performance metrics of emerging applications, and matches it to the statistical attributes of the underlying device and circuit fabrics [2]. We can trace the origins of stochastic computing to the works of John von Neumann. Stochastic computing circuits are able to realize arithmetic functions with very few logic gates. This is achieved by encoding numerical values within the statistics of random (or pseudorandom) binary sequences. Stochastic computing is an emerging computation manner which is considered to be promising for efficient probability computing. Compared with the conventional computation approach with deterministic manner, stochastic computing has smaller hardware footprint and higher energy efficiency because the basic multiplication and accumulation operations can be executed by AND gate and multiplexer (MUX), respectively. As a result, very large-scale integration on one chip is much easier to realize using stochastic computing than conventional fixed-point hardware [3].

There are many applications introduced for stochastic computing over the years. A particularly interesting application of stochastic computing is in the field of error correction decoding [4]. Another application of Stochastic computing is in Image processing. Image processing tasks require complicated arithmetic operations performed in parallel across every pixel in an image. Stochastic computing circuits bring the potential to implement these tasks with very low hardware cost [3]. Application of stochastic computing in Artificial Neural Networks led to the invention of Spiking Neural Networks (SNNs), which their objective is to model the stochastic activity of biological neurons. With the help of stochastic computing and stochastic neurons and their corresponding systems, we can achieve Brainware Large-Scale Integration. These BLSI systems are designed to simulate and compute certain functions as the brain does. Furthermore, with the combination of stochastic computing and advancement in device technology, we can design Neuromorphic chips. Designing Neuromorphic chips creates the basis of implementing BLSI systems and NNs on these chips to achieve extreme processing powers and higher energy efficiency.

The outline of the paper is as follows. Section II introduces an Unpredictive Non-determinism view of our surrounding world. After reviewing the stochastic computing from a data science basis, we proceed to section III. In this section, we compared the conventional computing method known as deterministic computing with stochastic computing. Also, we introduced different components and features of stochastic systems. Section IV summarizes the most common techniques and methods in implementing stochastic systems. These methods namely are: Algorithmic Noise Tolerance (ANT), Stochastic Sensor Network on Chip (SSNoC), Stochastic BitStream Computing, and Brain-inspired computing. The main purpose of section V is to introduce methods for optimizing stochastic systems in terms of latency. In section VI we talked about the implication on device design for stochastic computing. Also, we briefly reviewed the Invertible logics. Section VII is dedicated to the potential applications of stochastic computing. And finally, section VIII concludes this paper with insights and future directions.

## II. Stochastic Thinking

Uncertainty is one of the important concepts and foundations of our world. We face uncertainty in many areas on a daily basis. One of these areas is in Computer Sciences, or data science, or data computation. We rather deal with certainty and certain phenomena instead of uncertainty. For example, we rather have predictable functions, meaning that by giving input, we get the same output every time. It is not useful to treat any problem or phenomena as a certain issue and try to solve it by causal determinism. Nowadays, we are trying to raise the computational hierarchy, to help us better understand the world. And this goal is not attainable unless we consider uncertainty.

Whether or not the world is inherently unpredictable, the fact that we never have the complete knowledge of the world suggests that we might as well treat it as inherently unpredictable. We may refer to this statement as Unpredictive Non-Determinism.

In the following, we define a few concepts to help better understand the subject.

- ***Stochastic Process:*** An ongoing process when the next state might depend on both the previous states and some random elements.

The problem in simulating random events and calculating their probabilities in order to understand the effects of random events in a stochastic process is that today's computers are incapable of generating truly random numbers and data, they use algorithms to generate pseudorandom numbers. Therefore, these data are not random in nature.

- ***Simulation Models:*** A description of computations that provide useful information about the possible behaviors of the system being modeled.

When we say possible behaviors, we are particularly interested in stochastic behavior and stochastic systems.

Simulation Models are descriptive in nature, not prescriptive, in a sense that they describe the possible outcomes, but they don't tell us how to achieve a prospective outcome.

- ***Optimization Models:*** A prescriptive view of the modeled system. These optimization models tell us how to achieve an effect. For example, how to find the shortest path from A to B.

In contrast with optimization models, the simulation model shows that if we take a certain step, what will happen next as a consequence. But it doesn't tell us how to make something happen. Regardless of how these two models work, they both are essential in designing stochastic systems.

Now it is time to answer the question of what is probabilistic computing?

Advancements and evolution of technologies in modern days, the ever-growing volume of data, and user demands, all points in the direction the computers need to interpret and act on all of the data.

To fulfill this goal, we need computer systems capable of inductive inference, meaning they need to observe their world back to their underlying causes. Unfortunately, most of the computers that we use today are not suitable for these purposes because they are designed to solve scientific and specific problems. With this foundation, they cannot make sense of data. In a simple word, traditional computers execute a set of instructions that dictates them on how to transform inputs into outputs (how to map the input data to the corresponding outputs).

Typically, there are two ways that computers are used to interpret or understand data: Through Simulation, and Inference.

- In the simulation, the machine starts with some background assumptions and takes inputs as a configuration of the world, and produces an output and an observed trajectory. In this method, the computer executes some predefined/preprogrammed instructions in the same direction (from causes/inputs to effects/outputs).

-The inference is the reverse problem. In this method, the machine has the same background assumptions but takes as input an observed trajectory, and produces as output a configuration of the world that explains it. In this method, the direction is from the facts to their probable causes. A common problem with inference is that there are many plausible explanations/configurations for a particular output. There are uncertainty and unpredictability about which configuration/path is correct. Because of this unpredictable nature, we cannot expect certain answers, but we can aim for good guesses and achieve acceptable predictions about the data, which leads to less unnecessary complex configurations.

- ***probabilistic computing:*** based on the concepts introduced previously, probabilistic computing is all about managing and quantifying uncertainty about causal explanations.

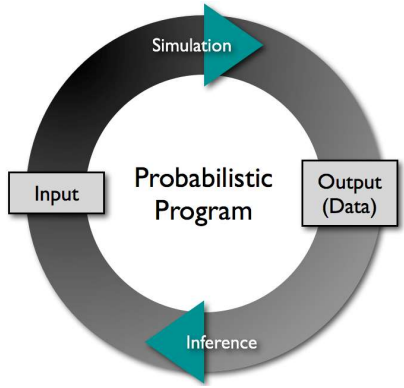

(a)

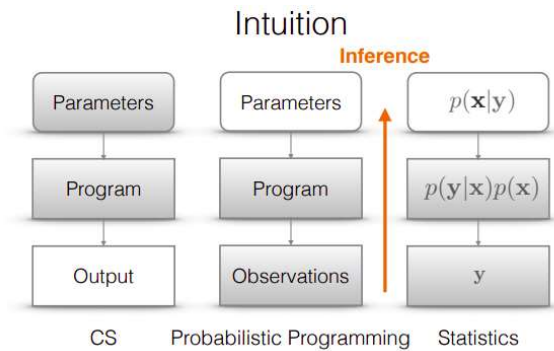

(b)

Figure 1. a) and b) both show how probabilistic computing works. program is the same as background assumptions, and is the same for both simulation and inference problems. But in the inference problem observations are the input of system.

### III. Fundamentals of Stochastic Computing [5], [6]

Energy efficiency and system reliability are the most important issues in the engineering world. All of the system engineers, device scientists, and experts in the field are working towards designing the next generation communication and processing technologies with maximum energy efficiency and reliability.

Designing energy-efficient devices will lead to energy-efficient systems. On the other hand, reliability at the system level does not mean reliability at device level structure, because we can build reliable systems out of unreliable devices.

At the system level, there is a very close tie end between energy efficiency and reliability. In other words, there is a tradeoff between energy efficiency and reliability. Stochastic computing takes the view that, if we need to achieve both energy efficiency and reliability, one needs to play around with the relationship between energy efficiency and reliability at all levels of the design hierarchy.

Foundations of computing are deterministic in nature. Most of the machines and computers that we are using today are designed based on Turing's deterministic finite state machine. This design architecture is based on Von Neumann's architecture. Therefore, computers and machines are built to handle deterministic computations. By looking at the origins of the device properties, we can observe that these devices have stochastical physics and origins. In deterministic computations, we hide and suppress this stochastic nature, to artificially represent ideal switch for the rest of the design hierarchy. At the higher levels (The user interface level), the level that the user is going to be utilizing these computing platforms, application metrics are also stochastic in nature. This is why designing the energy-efficient systems is such a hard problem, and nowadays we have an issue called power wall in the semiconductor and IC industry. Meaning, that lowering energy consumption is very hard and challenging at this point, and other methods need to be studied for designing new devices.

To tackle the deterministic computation's issues and satisfy the needs of industry for more energy-efficient and reliable devices, we need to start with Shannon's foundations, instead of basing the computational algorithm on the Turing machine and Von Neumann architecture. Shannon showed that by using statistical estimation and detection techniques, one could transmit information over a noisy channel with an arbitrarily small probability of error (Shannon – Hartley Theorem).

Stochastic computing says we shall view computing as a problem of transmitting information over a noisy channel, and use estimation and detection (statistical inference) to compensate for these kinds of errors. With this new point of view, we can conclude that there is no need to have ideal and perfect switches anymore. Meaning they can have Non-Determinism characteristics as well. Hopefully, this non-determinism results in tremendous energy savings. This new algorithm was extreme progress in system-level architecture because stochastic computation delivers the same results as deterministic computation with much less energy consumption, and users won't distinguish the difference in higher-level applications.

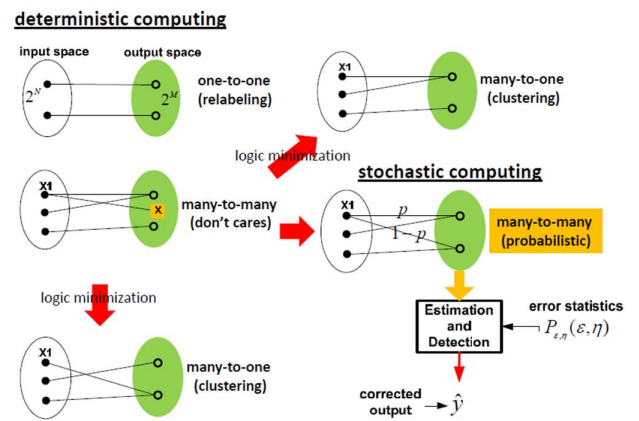

Figure. 2: Different configurations of Deterministic computing, and transforming metrics from Deterministic computing to Stochastic computing.

Von Neumann also observed that the treatment of error (in deterministic computations) is unsatisfactory, and therefore, the error should be treated as information has been [7].

To draw a path from Deterministic computing to Stochastic computing, we first need to categorize deterministic computing. We can categorize deterministic computing into three categories: 1) one-to-one. 2) many-to-one. 3) many-to-many.

Fig.2 better summarizes these configurations.

- **One – to – one:** It doesn't matter how many inputs and outputs we have (number of inputs and outputs might be equal or not). There is only one connection for each input and output. Meaning, one input is only connected to one output, and vice versa.

- **Many - to – one (clustering):** In this category, we can have more inputs connecting to one output. But the opposite is not true. Meaning one input can only connect to one output (multiple outgoing connections from one input is forbidden).

- **Many – to – Many (Don't cares):** In this category, one input may connect to more than one output. In other words, it has don't care state in the system.

Logic minimization dictates to map many-to-many configuration into many-to-one, by choosing the logic that leads to the smallest configuration or smallest logic netlist (by eliminating don't cares or choosing between one of the states to simplify the design complexity).

On the other hand, Stochastic computing can only be configured as probabilistic many-to-many.

- **Probabilistic Many – to – many:** As well as Deterministic many-to-many config, in this configuration, we can connect one input to more than one output (two outputs). Each connection has a probability value corresponding to that connection. The other outgoing connections from the same node, have complementary probabilities.

In stochastic problems, the problem of reliability boils down to taking the outputs and using error statistics, and by using estimation and detection techniques, we generate the final output (expected output).

Components of stochastic computing:

- **Error statistics:** One of the methods used in stochastic programing to calculate error statistics is Voltage over Scaling (VOS). Because we are dealing with modern-day processes, the way inducing error is by reducing the voltage level but keeping the clock frequency fixed.

This method causes lower power consumption for the same throughput. Also, this method allows us to observe output logic error, and finally calculate the probability distribution of errors. Since there is a trade-off between energy efficiency and reliability, we should set the power supply to the fixed voltage level that better compromises energy efficiency and effective error rate.

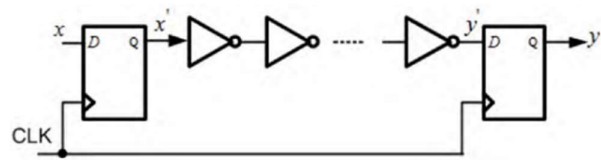

Figure.3 Simple block to understand VOS technique. Same clock frequency and different voltage levels, cause different probability distribution of error.

- **Statistical Estimation and Detection:**

Estimation and Detection fall into two major classes of techniques, in which we are always given a set of observations.

In the estimation problem, we are looking at outputs, and we want to figure out what is the correct output that generated these. In the ideal situation, we only have correct output values, but because of errors, these output values are obviating and migrating from correct values. Based on the number of observations, we want to estimate the correct output.

On the other hand, the detection problem uses the reversed approach. Meaning we have an Idea (Hypothesis) that the correct output belongs to a finite set. For example, between a few choices, we use observation to determine the choice of corresponding observation. In other words, we are trying to figure out that observation belongs to which case of the finite set.

Fig. 4 shows the framework of statistical Estimation and Detection Block.

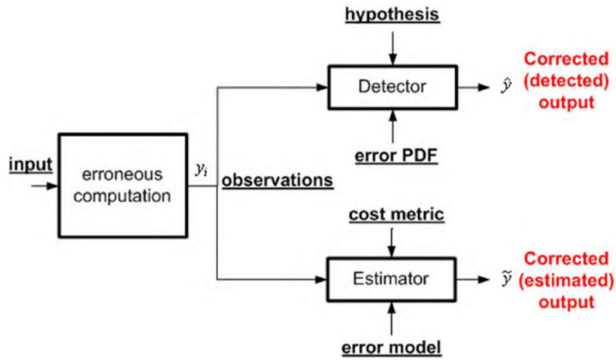

Figure. 4: Basic Frame work of Statistical Estimation and Detection block.

Stochastic computing is about taking application-level metrics and matching it to the statistics of the nanoscale fabrics.

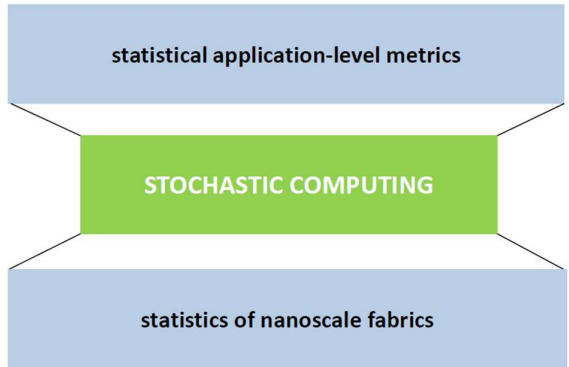

Figure. 5

One of the many features of the stochastic systems is skew tolerance.

- **Skew Tolerance:** If we have a delay in our system, in other words, if we are suffering from different arriving times of input data, the correct value is computed even when the inputs are misaligned temporarily [8].

Fig. 6 shows a multiplication operation on two input data with different arriving times.

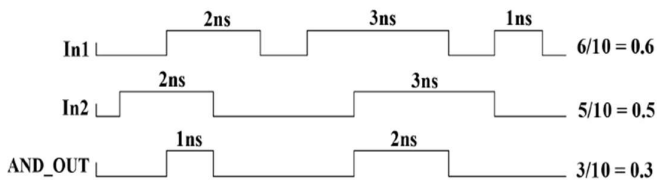

Figure. 6: stochastic multiplication using an AND with unsynchronized bit stream.

## IV. Stochastic Computing Techniques

There have been lots of developments over the years on implementation methods of stochastic systems. In this paper, we aim to review four of these methods and techniques. 1) Algorithmic Noise Tolerance (ANT). 2) Stochastic Sensor Network on Chip (SSNOC). 3) Stochastic Bitstream Computing. 4) Spintronic Approach for Stochastic Solutions.

In the following, we summarize each of these methods one-by-one [2].

- **Algorithmic Noise Tolerance:** There are two main blocks in ANT, which determine the function of ANT systems. After considering statistical concepts and fundamentals of stochastic computing, we let the main block to produce errors and operate under noisy conditions. This method will allow us to operate block at very high levels of energy efficiency. In contrast with the main block, the estimator should have lower complexity, which will lead to smaller circuitry and area sizes. Also, errors are not welcome in the estimator block. In other words, the estimator only should produce small errors, and not interfere with the main block's large errors [9].

Fig. 7 shows the basic block of ANT systems.

The estimator is a low-complexity computational block generating a statistical estimate of the correct main PE output, i.e.,

$$ya = yo + \eta$$
$$ye = yo + e$$

where $ya$ is the actual main block output, $yo$ is the error-free main block output, $\eta$ is the hardware error, $ye$ is the estimator output, and $e$ is the estimation error. The final/corrected output of an ANT-based system is obtained via the following decision rule:

$$\hat{y} = \begin{cases} y_a, & if\ |ya - ye| < \tau \\ y_e, & otherwise \end{cases}$$

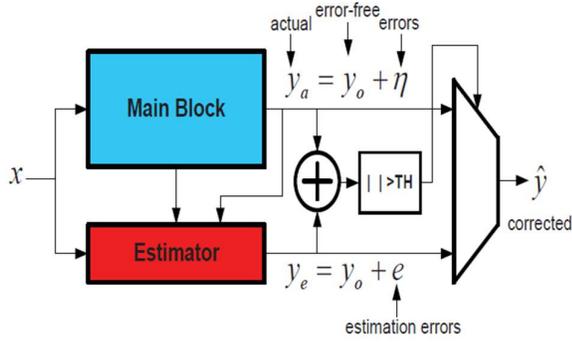

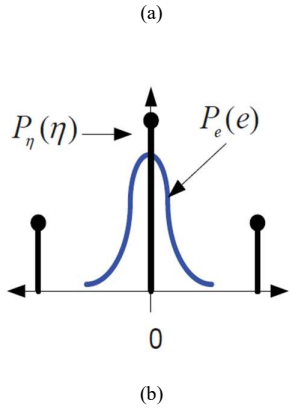

Figure. 7: a) framework of ANT system. b) error distributions.

- ***Stochastic Sensor Network on Chip (SSNoC):*** SSNOC relies only on multiple estimators or sensors to compute, permitting hardware errors to occur, and then fusing their outputs to generate the final corrected output [10].

Fig. 8 illustrates the basic structure of SSNoC system.

In SSNOC, we divide any computational block into subblocks. We call each of these subblocks, sensors. Each of the sensors and subblocks has both hardware errors and estimation errors. After dividing our computational function into subblocks, our designed system calculates the final output for each sensor based on Estimation and Detection techniques and the probability distribution of error. Finally, we fuse all of the outputs to generate the final corrected output [11].

The output of the *ith* sensor is given as:

$$y_{ei} = y_o + e_i + \eta_i$$

where $\eta_i$ and $e_i$ are the hardware and estimation errors in the *ith* estimator, respectively. Simulations indicate an $800\times$ improvement in detection probability while achieving up to 40% power savings.

In addition to ANT and SSNoC, there are few other communication-inspired stochastic techniques such as Soft NMR [12], Soft-Input Soft-Output computation [2], Stochastic Computation with Error statistics [2], and Likelihood Processing [5]. The effectiveness and general approach of these methods should be further studied during our research.

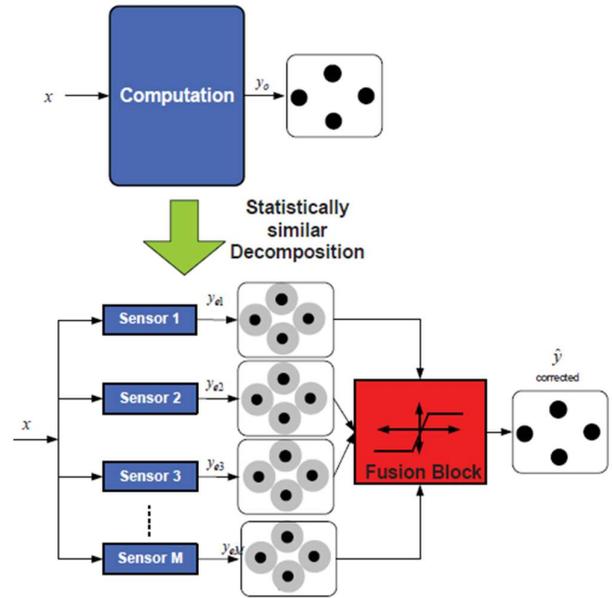

Figure. 8: framework of SSNoC system.

- ***Stochastic Bitstream Computing:*** Stochastic computing (SC) requires the generation of stochastic bitstreams: streams of randomly generated 1s and 0s, where the probability of an individual bit being 1 is $p$ and the probability of a bit being 0 is $1 - p$. Inputs, intermediate values, and outputs of stochastic circuits are represented with such bitstreams. Many approaches to stochastic computing require that input and intermediate value bitstreams be truly random, to the extent possible [3].

Fig.9 depicts a block diagram of a widely used circuit to generate a stochastic bitstream. A random number generator (RNG) produces a sequence of N-bit binary values: one such value per clock cycle. The random values are then fed into a comparator and compared with an N-bit binary number, B. Based on the comparison, a 0 or a 1 is generated in the stochastic bitstream [3].

To convert a stochastic bitstream back to a normal binary representation, a counter is typically used: on each clock cycle, when the stochastic stream contains a 1, the counter is incremented.

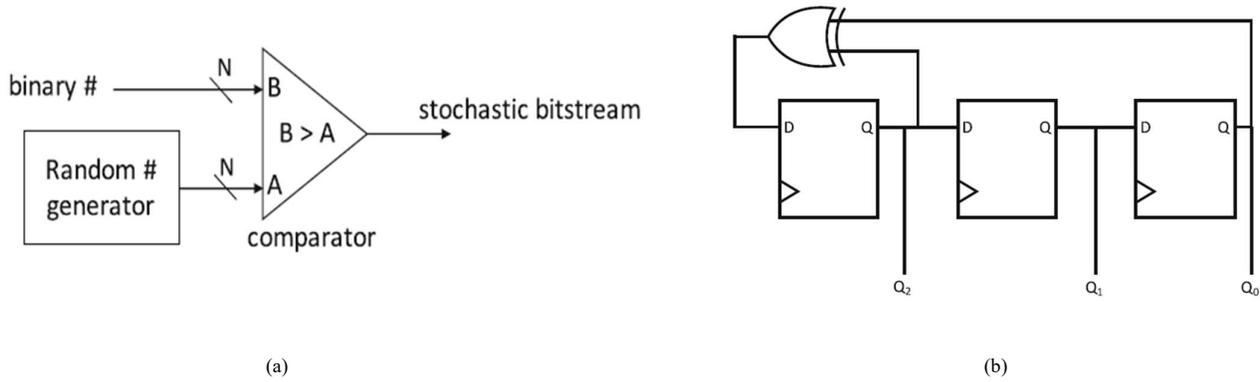

Figure. 9: (a) is a block diagram of a Random Number Generator (RNG). (b) block diagram of a 3-bit Linear feed-back shift register (LFSR).

With a length-L bitstream, after L clock cycles, if the counter holds value C, the value represented by the bitstream is C/L, which as expected lies in the [0 : 1] range.

This method is called Sequence generation, most widely used implementations for random number generators are Linear-feedback Shift register, and Low discrepancy Sequences [3].

- **Spintronic Approach for Stochastic Solutions:** Spintronic devices are utilized to realize efficient sampling operations to overcome the inference efficiencies in terms of power, area and speed. The intrinsic randomness existing in switching process of spintronic device is exploited to realize stochastic number generator. Stochastic computing is usually implemented by bit-wise operations with stochastic bitstreams which are generated by random number generator (RNG) and comparator. The stochastic switching of spintronic devices, such as magnetic tunnel junction (MTJ) provides a high-quality entropy source for RNG. Based on intrinsically unpredictable physical phenomenon, it can supply real random bitstreams by special circuit designs [3].

Based on the intrinsic stochastic behavior of the MTJ device, a TRNG (true random number generator) can be easily accomplished and used for stochastic number generator (SNG). A pair of transistors is sufficient to control the process of writing and reading [3].

## V. Optimization Methods for Stochastic Systems

So far, we have reviewed the fundamentals of stochastic systems and design techniques. Although stochastic computing offers simpler hardware for complex operations and has a higher noise tolerance than traditional deterministic systems, stochastic systems suffer from high latency and delay in the systems. In order to tackle this problem, we should consider different optimization methods.

As mentioned before, stochastic systems have Skew tolerance, which means they still can compute the correct output value even when the inputs have different arriving times.

Synchronism brings significant advantages to our systems, such as simplified design effort and guaranteed performance. But this advantage comes at significant costs. We need clock distribution networks (CDN) to synchronize our systems. Unfortunately, the CDN costs area consumes power and limits the system's performance.

-One study proposed a method entitled "Polysynchronous clocking" to tackle this problem and improve the high latency of stochastic systems[8]. This method is implantable with two different approaches:

1. Synchronize each domain using an inexpensive local clock. This method obviates the need for an expensive global CDN.

Or

2. Keep the global CDN but relax the clock skew requirements. This method allows for a higher working frequency.

Experimental results in [8] show that both of these Polysynchronous clocking methods improve latency in the stochastic systems significantly. In terms of latency, the first method (Removing local CDN) leads to much lower energy consumptions. In terms of area size, for large-scale systems, the first method (removing CDN) provides more area saving. However, for a smaller system second method (relaxing the clock) is the better approach [8].

-In serial communications and sequential computations, the receiver CPU must have a synchronized clock with the incoming data, in other words, the receiver CPU needs to have a synchronized clock with the sender CPU's clock, in order to have a fault-free and/or Latent-fault-free communication. One of the methods used for this purpose is called the Phase-Locked Loop (PLL). PLL is a closed-loop control system (a feedback system) that causes the generated signal at the receiver to stay synchronized with the reference signal. Analog PLLs require large on-chip capacitors whose leakage can seriously degrade PLL jitter performance. Therefore, Digital PLLs have few advantages over analog PLLs. For this purpose, in previous works, we designed and implemented ADPP (All-Digital Phase-Locked Loop).

Since we are dealing with stochastic computing, we need to use stochastic computing techniques to optimize ADPLL and use the Stochastic ADPLL for synchronizing our stochastic system. One study proposed a Digital Phase-Locked Loop with Calibrated Coarse and Stochastic Fine TDC [12].

The combination of proposed ADPLL and Stochastic Time digital converter (TDC) should be further studied. At this point, there is no guarantee for the effectiveness of this method in terms of improving latency. This technique requires further study.

## VI. Technology and Design

Since stochastic computing is an emerging area in designing energy-efficient and reliable systems, implementing them on device-level introduces new challenges on device structures and fabrication methods.

Important concepts on device design for stochastic computing for energy efficiency:

1. Non-deterministic device behavior.

We are used to non-determinism at system levels and considering the non-deterministic nature (physics) of devices, why we should force them to operate deterministically. In designing devices, we should embrace the non-deterministic behavior of devices.

2. Low SNR Switches:

In stochastic systems, it is acceptable If we are designing and fabricating switches that have a smaller gap between 1 and 0 logics (noisy 1s and 0s). We can use Estimation and Detection in the output and decode it to find out what is the correct output.

3. Multi-state switches:

In traditional switches, we have two states, on state and off state. The transition from one state to another consumes energy. In 2-state switches, because the energy gap between two steps is large, we have higher energy consumption. By utilizing multi-state switches, we can benefit from a closer energy gap between two different states, which leads to lower energy consumption in the transition from one state to another. Also, multi-state switches provide more design flexibility.

4. Invertible Logics:

Invertible logic has been recently presented for providing a capability of forward and backward operations as opposed to typical binary logic for the forward operation. It is designed based on underlying Boltzmann machines and probabilistic magneto-resistive device models (p-bits) [13].

Fig. 10 (a) shows a concept of invertible logic realized using Boltzmann machine and probabilistic bits (p-bits). Invertible logic circuits operate at forward and/or backward modes [14].

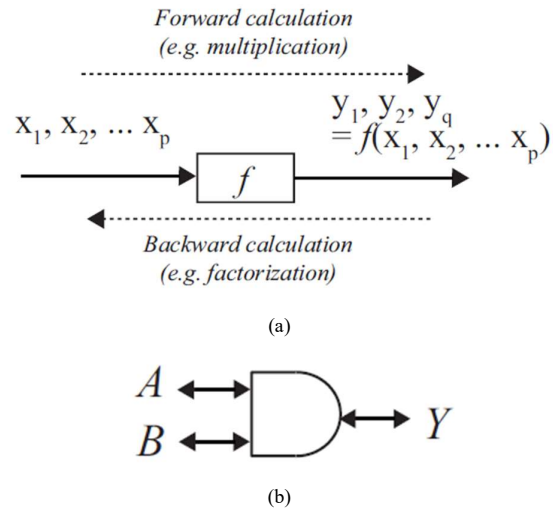

(a)

(b)

Figure. 10: a) concept of invertible logic, b) simple invertible AND

Designing Invertible logic is an important step in the realization of Stochastic systems. Designing and implementation of Invertible logics with CMOS and FinFET technology should be further studied.

Fig. 11 Shows realization of Hamiltonian Full Adder and 5-bit ReLU function.

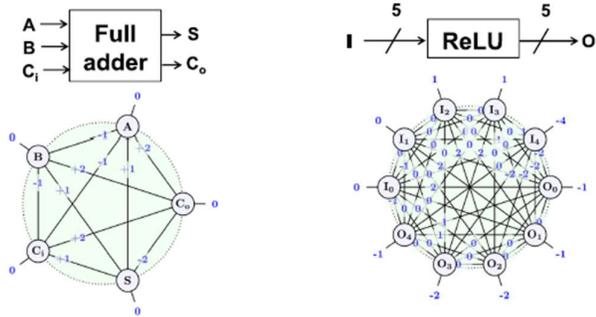

Figure. 11

## VII. Stochastic Computing Applications and Potential Research Areas

Now that we reviewed stochastic computing and its implementation techniques, it is time to point out the potential of Stochastic systems in emerging areas.

- ***Neuromorphic Computing:***

Designing and engineering computer chips that use the same physics of computation used by our brain's nervous system is called Neuromorphic computing.

This type of computing is fundamentally different from Artificial Neural Networks. ANN is a program ran on a conventional computer that mimics the logic of how the human brain thinks.

Neuromorphic Computing ⟶ Hardware version

Artificial Neural Networks ⟶ Software version

Neuromorphic chips and Artificial neural networks can work together because progress in both fields, especially Neuromorphic computing, will provide the possibility of running ANN on Neuromorphic hardware.

Traditional computers think in binary. They were designed by Von Neumann's architecture. In contrast, Neuromorphic computing works flexibly. Instead of using electrical signals to represent one or zero, designers of these new chips want to make their computer's neurons communicate with each other the way biological neurons do. Biological neurons use precise electrical current, which flows across a synapse (space between neurons).

This ability to transmit a gradient of understanding from neuron-to-neuron, and to have them all working together simultaneously, means that Neuromorphic chips could eventually be more energy efficient than our conventional computers, especially for complicated tasks. This is the place that stochastic computing could play an important role in realizing the neuron-to-neuron communications.

To realize this, we need new materials because what we are currently using in our computers is not enough. The physical property of silicon makes it harder to control the current flow between artificial neurons. There have been studies on single crystalline silicon and Tantalum Oxide to design devices with precise control over the current flow. University of Manchester's scientists developed and designed a Neuromorphic computing system based on traditional computing architecture. This system is called SpiNNaker, and they used traditional digital parts, like cores and routers, connecting and communicating with each other in innovative ways [15].

Neuromorphic computers offer the possibility of higher speed and more complexity for fewer energy costs.

Fig. 12 is the layout of IBM's Neuromorphic chip called "Truenorth" [16].

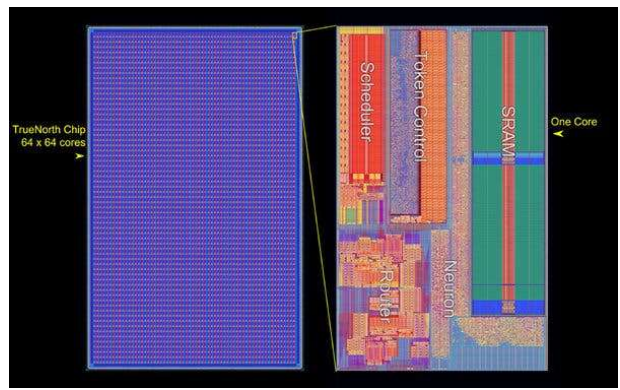

(a)

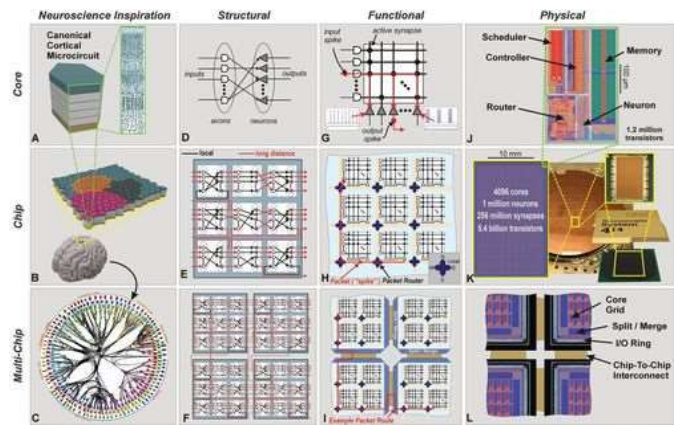

(b)

Figure. 12: a) layout of IBM's first Truenorth chip, b) Neuromorphic Computing based on IBM's Truenorth chip.

- *Brain-Inspired Computing:*

Brainware Large-Scale Integration (BLSI), is the result of brain-inspired computing based on stochastic computation. Stochastic computing exploits random bit streams, realizing area-efficient hardware for complicated functions such as multiplication and tanh, as compared with more traditional binary approaches.

Brainware computing requires complicated functions that can be area-efficiently realized using stochastic computing. The reason to choose stochastic computing for BLSI, is that human brains can perform well under severe noise and errors [3].

We can use BLSI systems to simulate or mimic the function of the human cortex. For example, in the designed BLSI system for implementation of Simple Cell of Primary Visual Cortex (2D Gabor Filter) [3].

Combination of Brainware Large-Scale Integration, Neuromorphic chips, and Stochastic computing is one of the interesting research areas that could provide possibilities for many future applications.

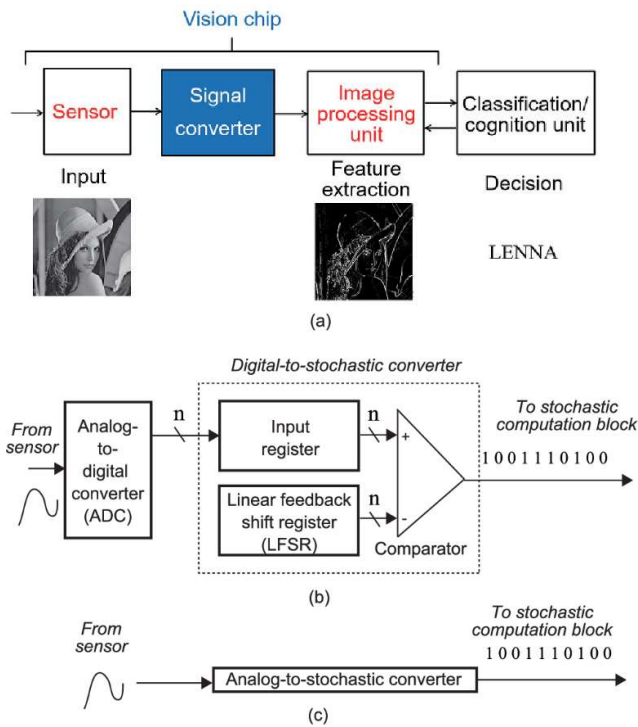

Figure. 13: Application of Stochastic computing in BLSI. a) Different blocks of Vision chip, b) Basic structure of Analog-to-Stochastic converter.

- *Stochastic Computing Devices and Invertible Logics:*

As we discussed before, Invertible logic has been recently presented for bidirectional operations using the Boltzmann machine and *p*-bits.

One of the encouraging research topics involves in designing more practical and more efficient invertible logic in terms of latency, energy consumption, and reliability. For future studies, different technologies such as CMOS, FinFET [17], and maybe Gate-All-Around FET (GAAFET), should be considered in engineering invertible logic.

- *Machine Learning:*

In recent years, Machine Learning and Deep Neural Networks (DNNs) have become the dominant approach for almost all recognition and detection tasks. The most common issues in designing and implementing Artificial Neural Networks (Convolutional NNs, Deep NNs, Recurrent NNs, and Spiking NNs) are energy consumption and hardware complexity. As the number of inputs and hidden layers increases in a Neural Network, Hardware complexity grows exponentially, and the designed hardware will occupy larger areas and will consume more power.

We have proven before that Stochastic Computing will provide a desirable tradeoff between energy efficiency and reliability in designing stochastic systems. Therefore, we can use Stochastic computing techniques for accelerating Artificial Neural Networks and achieve a better or even the same performance as traditional ANNs (ANN systems based on deterministic computing and Von Neumann's architecture) in terms of energy efficiency and reliability.

One study used stochastic computing for developing Image processing algorithms [18]. Fig. 14 compares the output of the designed image processing algorithm with Conventional Computing and Stochastic Computing methods.

VIII. Conclusion

In this paper, we briefly introduced Stochastic thinking from the Data science point of view. After considering a stochastic view of the world, we introduced fundamentals and basic concepts behind stochastic computing. After reviewing different implementation techniques, we can see that stochastic bit stream computing and spintronic approaches are the most promising methods in designing future stochastic systems.

Stochastic computing, as an emerging systems architecture, has the potential of breaking the power wall issue in the semiconductor and IC industry by designing energy-efficient and reliable devices. More research needs to be done in this area.

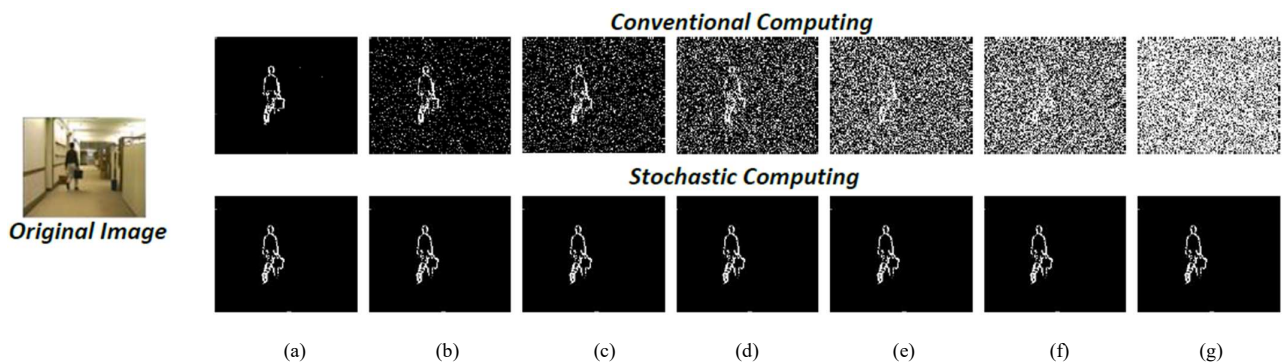

Figure. 14: A comparison of the fault tolerance capabilities of different hardware implementations of image processing algorithm. The images in the first row are generated by a conventional implementation. The images in the second row are generated using a stochastic implementation. Soft errors are injected at a rate of (a) 0%; (b) 1%; (c) 2%; (d) 5%; (e) 10%; (f) 15%; (g) 30%.

Hopefully, works and efforts in this area will lead to the state-of-art system architecture with the potential of achieving a life-long goal of implementing Neuromorphic chips with a processing power close to the human brain and designing more sophisticated BLSI systems.

For future work in this area, we could consider:

- Different Invertible Logic Designs (CMOS, FinFET, GAAFET)
- Designing BLSI systems focused on specific brain functions.
- Designing Neuromorphic Chips.